\documentclass[conference]{IEEEtran}

\usepackage{cite}
\usepackage[pdftex]{graphicx}
\graphicspath{{../pdf/}{../jpeg/}}
\DeclareGraphicsExtensions{.pdf,.jpeg,.png}

\usepackage{amsmath}
\usepackage{acronym}
\usepackage{amssymb}

\usepackage{array}

\usepackage{dblfloatfix}
\usepackage{url}

\usepackage{graphicx}
\graphicspath{ {./images/} }
\usepackage{wrapfig}
\usepackage{booktabs}

\begin{document}

\title{Estimating Mean Speed-of-Sound from Sequence-Dependent Geometric Disparities}

\author{Xenia~Augustin\textsuperscript{1},
        Lin~Zhang\textsuperscript{1},
        Orcun~Goksel\textsuperscript{1,2}\\
        \textit{\textsuperscript{1}Computer-Assisted Applications in Medicine, ETH Zurich, Switzerland} \\
        \textit{\textsuperscript{2}Department of Information Technology, Uppsala University, Sweden}}

\maketitle

\begin{abstract}
In ultrasound beamforming, focusing time delays are typically computed with a spatially constant speed-of-sound (SoS) assumption.
A mismatch between beamforming and true medium SoS then leads to aberration artifacts. 
Other imaging techniques such as spatially-resolved SoS reconstruction using tomographic techniques also rely on a good SoS estimate for initial beamforming.
In this work, we exploit spatially-varying geometric
disparities in the transmit and receive paths of multiple sequences for estimating a mean medium SoS.
We use images from diverging waves beamformed with an assumed SoS, and propose a model fitting method for estimating the SoS offset.
We demonstrate the effectiveness of our proposed method for tomographic SoS reconstruction.
With corrected beamforming SoS, the reconstruction accuracy on simulated data was improved by 63\% and 29\%, respectively, for an initial SoS over- and under-estimation of 1.5\%.
We further demonstrate our proposed method on a breast phantom, indicating substantial improvement in contrast-to-noise ratio for local SoS mapping.
\end{abstract}


\section{Introduction}
Speed-of-sound (SoS) is the longitudinal travel rate of sound in tissue.
Although SoS may change up to 5\% among tissues, ultrasound (US) image formation typically considers a spatially-constant SoS value in beamforming.
Thus, discrepancy between this assumed value and actual SoS lead to \emph{aberration} artifacts, for the correction of which SoS estimation was studied by several groups~\cite{shen2020ultrasound, imbault2017robust}, either directly correcting aberrations, e.g.\ using time-delayed signal statistics, or to estimate the medium SoS optimizing such statistics.
Most such work investigate homogeneous SoS estimation for gross aberration corrections.
A least-squares fit of a 2$^{nd}$-order polynomial to the echo profile was presented in~\cite{Anderson_direct_98} for the direct estimation of mean medium SoS, which was extended in~\cite{Byram_method_12} to measure SoS in layered phantoms.
An average SoS estimation by exploiting signal coherence among multiple transmit and receive paths was demonstrated in~\cite{shen2020ultrasound} for multi-angle plane wave imaging.

For tissue with highly heterogeneous SoS distribution, spatial SoS estimates can be used to correct aberrations locally.
A frequency domain reconstruction based on steered plane-waves was proposed in~\cite{Jaeger_computed_15} for estimating local SoS maps.
Spatial domain reconstruction methods were later proposed using plane~\cite{sanabria2018spatial} or diverging~\cite{Rau1} waves.
Rau et al.~\cite{Rau2} demonstrated a local-SoS adaptive beamforming based on SoS estimates from tomographic reconstruction.
However, such SoS reconstructions use beamformed images and  thus in turn rely on a good estimate of initial beamforming SoS.

\begin{figure*}[t]
	\centering
	\includegraphics[width=0.95\textwidth]{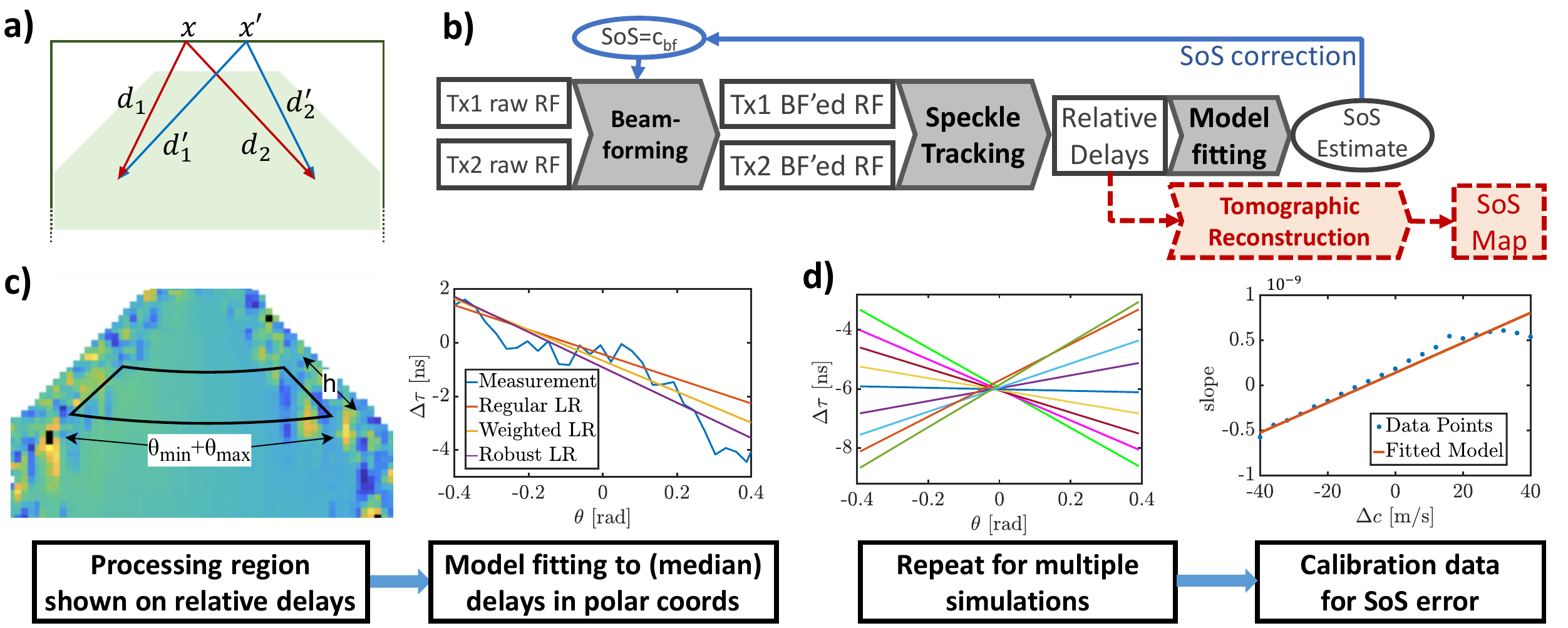}
	\caption{(a)~Illustration of geometric disparity with diverging waves. (b)~Processing pipeline for average SoS estimation (gray), its correction (blue), and its demonstrating for USCT (red). (c)~Model fitting to delays. (d)~Model calibration based on k-Wave simulations.}
	\label{fig:sosProcess}
\end{figure*}

In this work, we propose a mean SoS estimation method exploiting sequence-dependent geometric disparities.
After beamforming with any assumed SoS, differential measurements are taken using displacement estimation.
For any assumed SoS different from the true SoS, distance disparities cause an echo shift with a unique spatial pattern, which we fit with a model known from the transmission (Tx) and receive (Rx) geometry.
We demonstrate the effectiveness of this method in the context of the tomographic SoS reconstruction (USCT) with diverging wave transmits~\cite{Rau1}.
An overview of the processing pipeline is depicted in Fig.~\ref{fig:sosProcess}(b).

\section{Methods}
Considering two different diverging wave
transmits Tx1 and Tx2, with Rx apertures above each beamformed point, the image regions on the left and right as exemplified in Fig.~\ref{fig:sosProcess}(a) have differing wave travel distances $d(x)$ for a transceiver location $x$.
Any mismatch between the beamforming SoS $c_{\text{bf}}$ and the true medium SoS $c$ leads to spatially-varying echo shifts
\begin{equation}
    \Delta \tau_{\text{bf}} = \left(\frac{1}{c} - \frac{1}{c_{\text{bf}}}\right)d(x).
\end{equation}
The differential shifts between these two sequences, as illustrated in Fig.~\ref{fig:sosProcess}(a), can be written as
\begin{equation}
    \Delta \tau = \Delta \tau_{\text{bf}_1}-\Delta \tau_{\text{bf}_2}=\left(\frac{1}{c} - \frac{1}{c_{\text{bf}}}\right) \left( d(x)-d(x') \right).
\end{equation}
Echo shifts between two single element transmits is thus proportional to the difference between the inverse of $c$ and $c_{\text{bf}}$ as a function of the beamforming position.

In the following we demonstrate how to model the relation between the differential echo shifts and the mismatch between beamforming speed-of-sound (BF-SoS) and ground-truth medium speed-of-sound (GT-SoS).
To that end, we utilize the simulated data with known offset in BF-SoS.
An overview of our proposed correction method is illustrated in Fig.~\ref{fig:sosProcess}(c).

\subsection{Data generation}
To build and validate the correction method, we simulated a pulse-echo scenario with diverging waves following~\cite{Rau1}. 
The simulations were carried out using the \textit{k-Wave} toolbox~\cite{k-wave} in MATLAB. 
We used a single element transmission of a linear transducer with $N_c = 128$ channels and $300$\,mm pitch, with Tx pulses of $4$ half cycles with a center frequency $f_c = 5$\,MHz. 
The temporal resolution is $6.25$\,ns and the spatial discretization is $75$\,um per pixel.
The simulated raw channel data were beamformed using the delay-and-sum algorithm with different preset SoS values.
The relative delays between beamformed images are measured using a normalized cross correlation method along the axial direction as in~\cite{Rau1}. 
Without loss of generality, we used herein the time delay measurements between the frames from diverging transmits from the elements $55$ and $65$.
We used dynamic receive focusing with full aperture, so the Rx paths are the same between the frames and only the Tx paths differ.

\subsection{Geometric Disparity Patterns}
To model the relation between $\Delta c=c_{bf}-c$ and relative delay measurements $\Delta \tau$, we first investigate how $\Delta\tau$ changes spatially (laterally) in different cases of $\Delta c$.
To that end, we select a region of interest with a depth of $h$ in the delay measurements, as indicated in Fig.~\ref{fig:sosProcess}, to calculate the median relative delay values over the chosen depth.
Since the ultrasound diverging waves propagate with circular wave fronts, we choose to use polar coordinates.
We consider an angular range of $[-\theta_{\text{min}}, \theta_{\text{max}}]$.
An example plot of the median delay value against the angular coordinate is shown in Fig.~\ref{fig:sosProcess}(c).
From left to right, we observe a monotonically reducing delay pattern, if the BF-SoS is underestimated; and an increasing pattern, for vice versa.
These patterns are due to the spatially differing Tx paths, as also formalized in Eq\,(2).
Since a nearly-linear relationship is observed between relative delays (median values over a certain depth range) and the angle $\theta$, we herein use linear regression to quantify these delay patterns.

\subsection{Linear Regression}
For linear regression, we studied herein the standard least-squares, a robust fitting with \emph{iteratively-reweighted least squares}, and a weighted fitting approach.
For weighted linear regression~\cite{Poree}, we use the cross-correlation values of the relative delay measurements, which serve as a confidence measure, as the weighting factors; yielding 
\begin{equation}
	b = (X^T\cdot W \cdot X)^{-1} \cdot X^T \cdot y,
\end{equation}
where $b$ denotes the regression model parameters sought, $X$ contains the data points, $y$ the observed delay values, and $W$ is the diagonal weighting matrix.

\subsection{Calibration Model}
For each $\Delta c$, the above model regression then yields a slope value that represents the relation between the lateral position $\Theta$ and the observed time-delay $\Delta \tau$ pattern.
Using these observations, we then build a calibration model such that at application time, given a time-delay pattern observation from any BF-SoS, we can estimate the error in BF-SoS for correction.
For given (intentional) beamforming SoS errors $\Delta c$, we present the slopes calculated from time-delay observations in Fig.~\ref{fig:sosProcess}(d).
To create a calibration model, we fit a polynomial of degree $n$ (studied below) to these slope observations.
Using this calibration as a predictive model for inverse-lookup, for any observed time-delay pattern (slope), we can then find the BF-SoS error $\Delta c$, i.e.\ the difference of the assumed SoS to the true SoS.

\subsection{Image Reconstruction with BF-SoS Correction}
Reconstruction of slowness (inverse of SoS) map $\sigma 
\in \mathbb{R}^{N_xN_y}$, can be formulated as the following inverse problem
\begin{equation}
    \hat{\sigma} = \text{argmin}|| L\sigma - \Delta\tau ||_1 + \lambda||D\sigma||_1,
\end{equation}
where $L$ is the differential path matrix determined by acquistion geometry and $\Delta \tau \in \mathbb{R}^{MN'_xN'_z}$ is the vectorized time delay measurement with $M$ combinations of beamformed RF frames.
$D$ denotes the regularization matrix and $\lambda$ controls the regularization strength.
We use the total variation regularization with anisotropically weighted spatial gradients as in~\cite{Rau1}.
Following~\cite{Bernhardt} we use $6$ combinations of frames beamformed with the corrected BF-SoS value.
This optmization problem is solved using limited memory Broyden-Fletcher-Goldfarb-Shanno (L-BFGS) algorithm~\cite{Rau1}.

\section{Results}
\label{sec:results}
To compare regression models, we use goodness of fit
\begin{equation}
    R^2 = 1 - \frac{\sum_{i}(y_i - \hat{y}_i)^2}{\sum_{i}(y_i - \bar{y})^2}
\end{equation}  
with the $i$-th observed data point $y_i$, predicted $\hat{y}_i$, the mean value of all the observations $\bar{y}$, and the root mean squared error (RMSE) between the observed and fitted data.
For evaluating tomographic SoS reconstruction, we use RMSE between the ground truth and reconstructed SoS maps; as well as
contrast-to-noise-ratio $\text{CNR}=2(\mu_{inc}-\mu_{bkg})^2/(\sigma_{inc}^2+\sigma_{bkg}^2)$ with the mean SoS value of the inclusion $\mu_{inc}$ and background $\mu_{bkg}$ and the corresponding standard deviations $\sigma_{inc}$ and $\sigma_{bkg}$.
We report CNR on the decibel scale.
Since there is no ground truth label for the phantom images, we only compute CNR for the phantom evaluation.

\vspace{1ex}\noindent{\bf Linear regression. }
We herein select the delay measurements with a depth of $7.5$\,mm between $7.5$ and $15$\,mm and within the angular range of $[-0.4,0.4]$~rad.
We use the delays near the probe, which is assumed to be less influenced by any deeper tissue inhomogeneities, e.g.\ inclusions.
The $R^2$ measure is used to select the optimal depth and angular range.
We observed that averaging delay measurements over a larger image depth is more robust to outliers and noisy measurements around the aperture edges can be avoided using a smaller angular range.
Three different fitting algorithms: regular, robust, and weighted linear regressions are investigated.
The average RMSE and $R^2$ over all the simulated data are shown in Tab.~\ref{tab1}.
\begin{table}
\renewcommand{\arraystretch}{1.3}
\caption{Quantitative results of different regression models}
\label{tab1}
\centering
\begin{tabular}{p{1cm} p{1.6cm} p{1.6cm} p{1.6cm}}
             & \textbf{Regular LR} & \textbf{Robust LR} & \textbf{Weighted LR}\\ \toprule
			$R^2$ & 0.83 & 0.93  & 0.84 \\ 
			$\text{RMSE}$ & 0.23 & 0.24  & 0.27 \\
			\bottomrule
\end{tabular}
\end{table}
Compared to the regular and weighted fitting, the robust regression approach achieves the best fitting performance with a $R^2$ value of $0.93$. 
The regular fitting approach is more prone to outliers due to noisy delay measurements, whereas the cross correlation values used for the weighted fitting could contain noisy weighting information.

\begin{figure}
	\centering
	\includegraphics[width=0.3\textwidth]{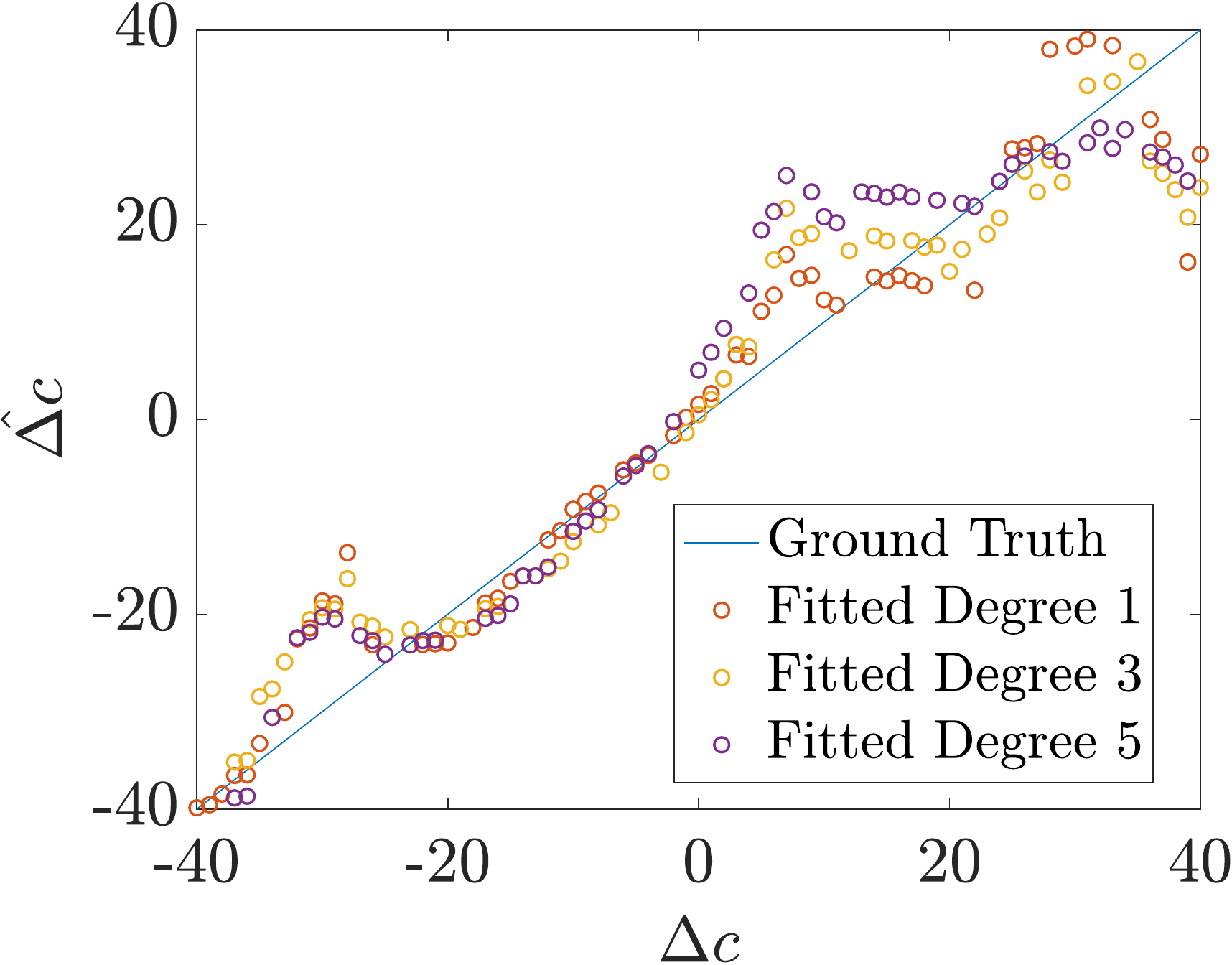}
	\caption{Scatter plot of the ground truth $\Delta c$ against the estimated SoS offset $\hat{\Delta c}$ using different polynomial fits}
	\label{fig:result2}
\end{figure}

\vspace{1ex}\noindent{\bf Calibration model. } 
We performed simulation for a homogeneous medium with a GT-SoS value of $1500$~m/s and beamformed with $\Delta c=c-c_{bf}$ ranging from $-40$~m/s to $+40$~m/s  with a grid size of $1$~m/s, leading to $81$ different beamformed images.
The calibration model was built using 21 equally spaced data points.
The remaining 60 data points were used for testing the calibration model for the polynomial fits of degrees 1, 3, and 5, by comparing the estimated SoS offsets to the known ground truth offset values, as shown in Fig.~\ref{fig:result2} and tabulated in Tab.~\ref{tabNew}.
Although RMSE is the lowest for the slope model with the polynomial degree of~5, herein we choose to use degree~1 to prevent over-fitting.
We observed similar results while using simulated images with other numerical ground truth SoS phantoms.

\begin{table}
\renewcommand{\arraystretch}{1.3}
\caption{Quantitative results for the calibration model using different polynomial degrees}
\label{tabNew}
\centering
\begin{tabular}{p{1cm} p{1.6cm} p{1.6cm} p{1.6cm}}
             & \textbf{Degree 1} & \textbf{Degree 3} & \textbf{ Degree 5}\\ \toprule
            $R^2$ & 0.94 & 0.93 & 0.94 \\
			RMSE & 3.67 & 3.27  & 2.17 \\
			\bottomrule
\end{tabular}
\end{table}

\begin{figure*}
	\centering
    \includegraphics[width=0.95\textwidth]{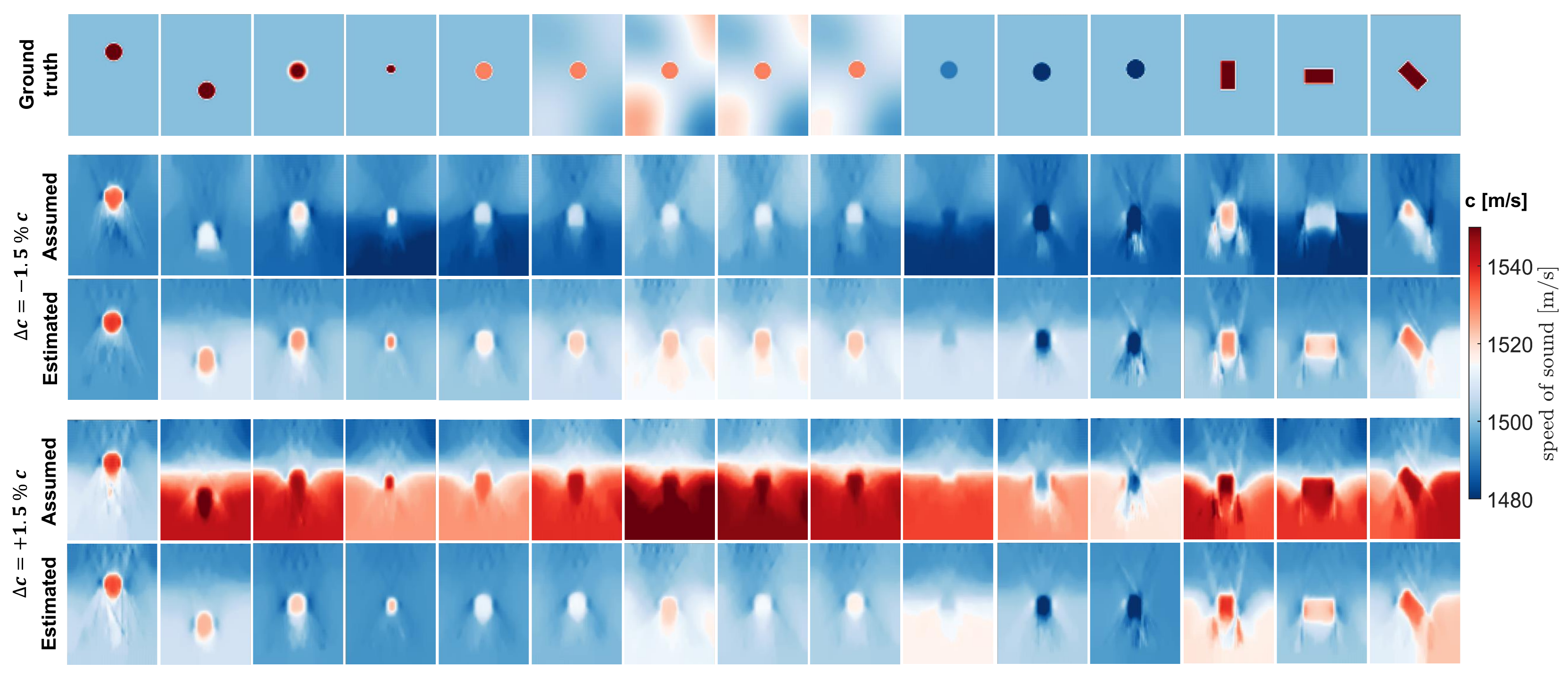}
	\caption{SoS reconstruction results on k-Wave simulated images in case of SoS underestimation (top) and overestimation (bottom).}
	\label{fig:result3}
\end{figure*}

\vspace{1ex}\noindent{\bf BF-SoS correction on simulated data. } 
We evaluate the effectiveness of the proposed calibration model for the SoS reconstruction on the simulated data.
We use 32 simulated test images reported in~\cite{Bernhardt} including elliptical and rectangular inclusions with varying inclusion sizes, positions, and background SoS variations.
We beamformed the raw RF channel data with two SoS values, which are $1.5 \%$ larger ($\Delta c_1$) and $1.5\%$ smaller ($\Delta c_2$) than the mean ground truth SoS, to simulate the over- and under-estimation case.
For these cases, our proposed BF-SoS correction method respectively reduces the reconstruction RMSE by $63 \%$ and $29 \%$, as tabulated in Tab.~\ref{tab:RMSEReco}.
\begin{table}    
    \caption{RMSE and CNR between the ground truth and the reconstructed SoS maps before and after BF-SoS correction by our proposed method}
    	\begin{center}
    		\begin{tabular}{lrrrr}\toprule				
    			& \multicolumn{2}{c}{\textbf{RMSE}} &\multicolumn{2}{c}{\textbf{CNR}} \\ \cmidrule{2-5}
    			& \multicolumn{1}{c}{\textbf{before}} &\multicolumn{1}{c}{\textbf{after}} & \multicolumn{1}{c}{\textbf{before}} &\multicolumn{1}{c}{\textbf{after}}\\
    			\cmidrule{2-5}
    			$\Delta~c_1 $ & 24.70 & 9.11 & 0.93 & 8.65 \\
     			$\Delta~c_2 $ & 11.78 & 8.35 & 10.35 & 8.02 \\  			
    			\bottomrule
    		\end{tabular}
    	\end{center}

    	\label{tab:RMSEReco}
\end{table}
Examples of simulated images are shown in Fig.~\ref{fig:result3}.

\begin{figure}
	\centering
	\includegraphics[width=0.95\linewidth]{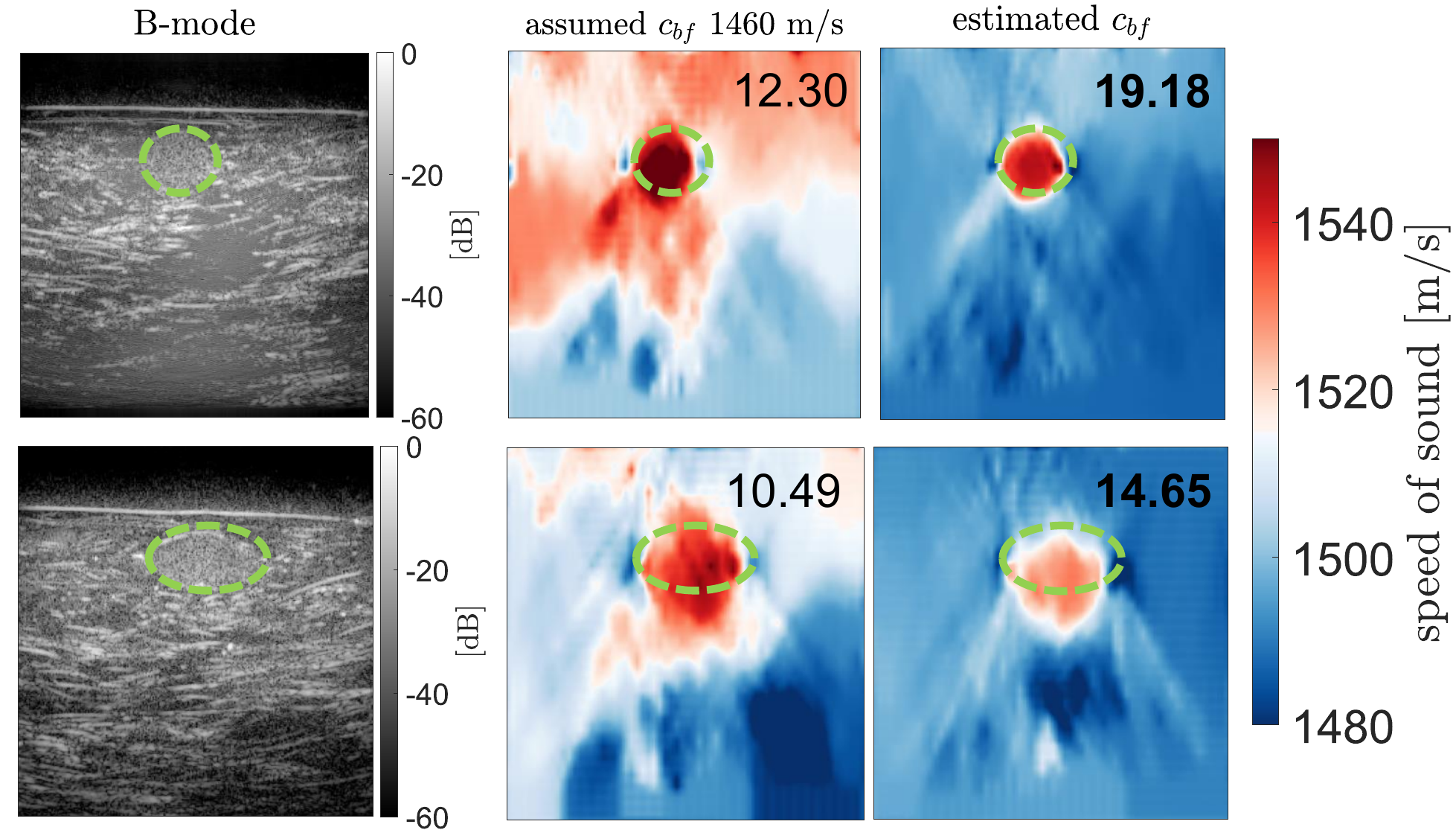}
	\caption{SoS reconstruction for the two phantom examples with (left) the Bmode images, (center) the reconstructed SoS maps with the initial assumed BF-SoS of 1460\,m/s, and (right) the corrected BF-SoS by our method as 1491 m/s (top) and 1494 m/s (bottom). CNR values are reported as inset.}
	\label{fig:result5}
\end{figure}

\vspace{1ex}\noindent{\bf BF-SoS correction for phantom. }
The proposed correction model is further evaluated on a breast phantom, with data collected using a Fukuda Denshi UF-760AG with the linear probe FUT-LA385-12P.
For transmits with a center frequency of $5$~MHz and 4 half cycles, we collected pre-beamformed raw channel data.
Fig.~\ref{fig:result5} shows qualitative reconstruction results for an initial BF-SoS estimate of 1460 m/s, and those for BF-SoS corrected with our method.
Reconstructions are seen to be largely improved, as also indicated by the CNR improvements.

\section{Conclusion}
In this work we have demonstrated a beamforming speed-of-sound correction method for ultrasound imaging. 
We utilize relative delay measurements between image frames with diverging transmits, and show that the spatial patterns of such delays reflect the mismatch between the true and the beamforming speed-of-sound.
To model the relationship between the measured delays and the offset in BF-SoS, an angular coordinate is linearly regressed to delays averaged over a selected region of interest.
For a robust data-based model we employ a calibration stage using k-Wave simulations.
The proposed correction method leads to an improvement in SoS reconstruction on the simulated images by $63$\,\% and $29$\,\% in RMSE in case of over- and under-estimation, respectively.
Further evaluation on breast phantom data shows the feasibility of our proposed method with a physical imaging setup.

\bibliographystyle{IEEEtran}
\bibliography{refs}

\end{document}